\documentclass{aa}  

\usepackage{graphicx}
\usepackage{caption}
\usepackage{subcaption}
\usepackage{txfonts}
\usepackage{gensymb}
\begin{document}

   \title{The similarity of the interstellar comet 2I/Borisov to solar system comets from high resolution optical spectroscopy}

   \subtitle{}
\titlerunning{}

\author{
C. Opitom\thanks{E-mail: copi@roe.ac.uk} \inst{1}
\and
E. Jehin \inst{2}
\and
D. Hutsemékers \inst{2}
\and
Y. Shinnaka \inst{3}
\and
J. Manfroid \inst{2}
\and 
P. Rousselot \inst{4}
\and
S. Raghuram \inst{5}
\and
H. Kawakita \inst{3,6}
\and
A. Fitzsimmons \inst{7}
\and
K. Meech \inst{8}
\and
M. Micheli \inst{9}
\and
C. Snodgrass \inst{1}
\and
B. Yang \inst{10}
\and
O. Hainaut \inst{11}
}

\institute{
Institute for Astronomy, University of Edinburgh, Royal Observatory, Edinburgh EH9 3HJ, UK
\and
Space sciences, Technologies \& Astrophysics Research (STAR) Institute, University of Liège, Liège, Belgium
\and 
Koyama Astronomical Observatory, Kyoto Sangyo University, Motoyama, Kamigamo, Kita-ku, Kyoto 603- 8555, Japan
\and
Institut UTINAM UMR 6213, CNRS, Univ. Bourgogne Franche-Comté, OSU THETA, BP 1615, 25010 Besançon Cedex, France
\and
Laboratory for Atmospheric and Space Physics, University of Colorado Boulder, Boulder, CO, USA
\and
Department of Astrophysics and Atmospheric Sciences, Faculty of Science, Kyoto Sangyo University, Motoyama, Kamigamo, Kita-ku, Kyoto 603-8555, Japan
\and
Astrophysics Research Centre, School of Mathematics and Physics, Queens University Belfast, Belfast BT7 1NN, UK
\and
Institute for Astronomy, 2680 Woodlawn Drive, Honolulu, HI 96822 USA
\and
ESA NEO Coordination Centre, Largo Galileo Galilei, 1, 00044 Frascati (RM), Italy
\and
European Southern Observatory, Alonso de Cordova 3107, Vitacura, Santiago, Chile
\and
European Southern Observatory, Karl-Schwarzschild-Strasse 2, D-85748 Garching bei München, Germany
}

   \date{}

  \abstract{}{2I/Borisov - hereafter 2I - is the first visibly active interstellar comet observed in the solar system, allowing us for the first time to sample the composition of a building block from another system. We report on the monitoring of 2I with the Ultraviolet-Visual  Echelle  Spectrograph  (UVES), the high resolution optical spectrograph of the ESO Very Large Telescope at Paranal, during four months from November 15, 2019 to March 16, 2020. Our goal is to characterize the activity and composition of 2I with respect to solar system comets.} {We collected high resolution spectra at 12 different epochs from 2.1 au pre-perihelion to 2.6 au post perihelion.}{On December 24 and 26, 2019, close to perihelion, we detected several OH lines of the 309~nm (0-0) band and derived a water production rate of $2.2\pm0.2 \times 10^{26}$ molecules/s. The three [OI] forbidden oxygen lines were detected at different epochs and we derive a green-to-red doublet intensity ratio (G/R) of $0.31\pm0.05$ close to perihelion. NH$_2$ ortho and para lines from various bands were measured and allowed us to derive an ortho-to-para ratio (OPR) of $3.21\pm0.15$, corresponding to an OPR and spin temperature of ammonia of $1.11\pm0.08$ and $31^{+10}_{-5}$ K, respectively. These values are consistent with the values usually measured for solar system comets. Emission lines of the radicals NH (336~nm), CN (388~nm), CH (431~nm), and C$_2$ (517~nm) were also detected. Several FeI and NiI lines were identified and their intensities were measured to provide a ratio of log (NiI/FeI) = $0.21\pm0.18$ in agreement with the value recently found in solar system comets.}{Our high spectral resolution observations of 2I/Borisov and the associated measurements of the NH$_2$ OPR and the Ni/Fe abundance ratio are remarkably similar to solar system comets. Only the G/R ratio is unusually high but consistent with the high abundance ratio of CO/H$_2$O found by other investigators.}


   \maketitle
%

\section{Introduction}

Interstellar objects provide an unprecedented opportunity to probe material formed under potentially very different conditions from solar system comets. 2I/Borisov (hereafter 2I) is only the second interstellar object discovered and the first for which activity was directly detected. 2I was discovered in August 2019 at about 3 au from the Sun, prior to its perihelion passage, and was already active at the time of discovery. 2I remained observable with major astronomical facilities for several months. This provided a unique opportunity to measure the composition of its coma and probe the ice composition of a comet formed around another star. 

Early observations, mainly with optical low-resolution spectrographs, showed that 2I is depleted in C$_2$, similarly to solar system carbon-chain depleted comets (\cite{Fitzsimmons2019,Opitom2019,Bannister2020,Kareta2020,Lin2020,Aravind2021}). Later, \cite{Cordiner2020} and \cite{Bodewits2020} reported that 2I is very rich in CO relative to both water and HCN, more than any solar system comet observed within 2.5 au from the Sun. This provided the first indication that the composition of the first active interstellar comet might differ from that of most solar system comets. Recent results indicated changes in the mixing ratio of species in the coma of 2I between pre- and post-perihelion observations, hinting at an heterogeneous nucleus \citep{Xing2020,Yang2021,Aravind2021}. 

We present here high-resolution spectroscopic observations of 2I performed at the Very Large Telescope, aimed at constraining the composition of the interstellar comet. In this work, we focus mainly on the gas component of the coma.

\section{Observation and data reduction}
We obtained 15 spectra of 2I on 12 dates (for a total of $\sim21$~hours) with the Ultraviolet-Visual Echelle Spectrograph (UVES) mounted on the ESO 8.2 m Very Large Telescope. We used 3 different settings to cover the full optical range: the dichroic\#1 (346+580) setting covering the range 303 to 388 nm in the blue and 476 to 684 nm in the red; the free setting (348+580) which is very similar to the (346+580) setting, but for which the blue part is shifted by about two nm to the red; and the dichroic \#2 (437+860) setting covering the range 373 to 499 nm in the blue and 660 to 1060 nm in the red. UVES has a slit length of 10\arcsec$ $ and we used slit widths of 0.4\arcsec, 0.7\arcsec, and 1.8\arcsec, providing resolving powers of about $80\,000$, $50\,000$, and $30\,000$, respectively. The observations were spread over 4 months between Nov 15, 2019 and Marc 16, 2020. Most of the observations analysed here were near perihelion on Dec 8, 2019. Very long exposures of up to 2 h were used, to minimize the effect of the read-out noise for these observations of a faint target. The observing dates, set-ups, and observational circumstances are presented in Table \ref{TableObs}. 

The ESO UVES pipeline\footnote{ftp://ftp.eso.org/pub/dfs/pipelines/instruments/uves/uves-pipeline-manual-6.1.3.pdf} was used for the basic reduction, including wavelength calibration, extinction correction, and flux calibration using master response curves. We compared the flux calibration using master response curves and standard stars observed close to the science observations and found no significant differences. Since the UVES pipeline is not optimised for very extended objects, we used custom routines for the spectrum extraction over the full slit length and cosmic rays removal. The spectra were corrected for the Doppler shift due to the geocentric velocity of the comet. Finally, the dust continuum (as well as moon and twilight contribution when necessary) were removed using a BASS2000 solar spectrum re-sampled to match the resolution of the observations in each setting. The BASS2000 was fitted individually for each component (dust-reflected sunlight, twilight or moon contamination), allowing us to properly correct for the reddening of the comet continuum for example. More details concerning the data reduction can be found in \cite{Manfroid2009}.

\begin{table*}
\caption{\label{TableObs} Observing circumstances of the 2I VLT campaign}
\centering
\begin{tabular}{lcccccccc}
\hline\hline
Date (UT) & Instrument Set-up & Slit Width (\arcsec) & N & Exposure time (s) & r$_h$ (au) & $\mathrm{\dot{r}}_h$ (km/s) & $\Delta$ (au) & $\dot{\Delta}$ (km/s)\\
\hline
2019-11-15 & 348+580 & 0.7 & 1 & 3000 & 2.1 & -9.6 & 2.2 & -22.8 \\
2019-12-16 & 346+580 & 1.8 & 1 & 3000 & 2.0 & 3.3 & 2.0 & -6.4 \\
2019-12-24 & 346+580 & 1.8 & 1 & 6600 & 2.0 & 6.6 & 1.9 & -2.2 \\
2019-12-25 & 437+860 & 1.8 & 1 & 6600 & 2.0 & 7.0 & 1.9 & -1.7 \\
2019-12-26 & 346+580 & 1.8 & 1 & 6600 & 2.0 & 7.4 & 1.9 & -1.2 \\
2020-01-29 & 346+580 & 0.4-1.8 & 2 & 3350-6600 & 2.3 & 18.9 & 2.1 & 11.5 \\
2020-02-01 & 346+580 & 0.4 & 1 & 4500 & 2.3 & 19.7 & 2.1 & 12.2 \\
2020-02-02 & 346+580 & 0.4 & 1 & 4500 & 2.4 & 19.9 & 2.1 & 12.4 \\
2020-02-04 & 346+580/437+860 & 1.8 & 2 & 6600 & 2.4 & 20.4 & 2.1 & 12.9 \\
2020-02-05 & 346+580 & 1.8 & 1 & 3000 & 2.4 & 20.7 & 2.1 & 13.2 \\
2020-02-22 & 348+580 & 1.8 & 2 & 2870-4875 & 2.6 & 24.2 & 2.3 & 16.0 \\
2020-03-16 & 348+580 & 1.8 & 1 & 6600 & 2.6 & 24.2 & 2.3 & 16.0 \\
\hline
\end{tabular}
\tablefoot{N represents the number of exposures obtained}
\end{table*}

\section{Analysis and results}

2I, with a R-band magnitude around 16.5 at perihelion \citep{Jehin2020}, was relatively faint compared to comets usually observed at high spectroscopic resolution. In spite of that, emissions from several species were detected: OH, NH, CN, CH, C$_2$, NH$_2$, [OI], and metals FeI and NiI (see an extensive list in Table \ref{TableSpec}). Unfortunately, none of the detected species had sufficiently strong emission bands to measure isotopic ratios. We searched for but did not detect emissions from N$_2^+$, CO$^+$, CO$_2^+$, H$_2$O$^+$, and C$_3$.


\subsection{Nickel and iron abundances}
\label{FeNi}

\cite{Manfroid2021} recently showed, using UVES high resolution spectra collected over the last 20 years, that emission lines of neutral FeI and NiI atoms are ubiquitous in the atmospheres of a number of solar system comets of various composition, dynamical origin, and at different distances from the Sun (up to 3.2 au). NiI was also detected in the coma of 2I by \cite{Guzik2021}.
We searched our UVES spectra of 2I and identified several FeI and NiI lines (see Table \ref{TableSpec} and Fig. \ref{Plot_NiFeLines}). We followed the method described in \cite{Manfroid2021} and applied their multilevel atomic model (taking into account the true solar spectrum) to derive FeI and NiI production rates for 2I. We only used a subset of the lines listed in Table \ref{TableSpec} for this analysis, as some of these lines were of too low SNR or blended with other emission lines. The production rates were derived using an average spectrum from Dec 24 and 26, 2019 and a spectrum obtained on Jan 29, 2020, as these were the only spectra with sufficient signal. Due to the different heliocentric velocities of the comet, the production rates for the two epochs were computed separately and then averaged, providing values of log~Q(NiI)=~21.88 $\pm$ 0.07 molecules/s and log~Q(FeI)=~21.67 $\pm$ 0.16 molecules/s, corresponding to log(NiI/FeI)=~$0.21\pm0.18$. Our measurement of the NiI abundance is in very good agreement with the $9\pm3\times10^{21}$ molecules/s reported by \cite{Guzik2021} for observations performed on Jan 28, 30, and 31, 2020.

\begin{figure}
\centering
\includegraphics[width=9cm]{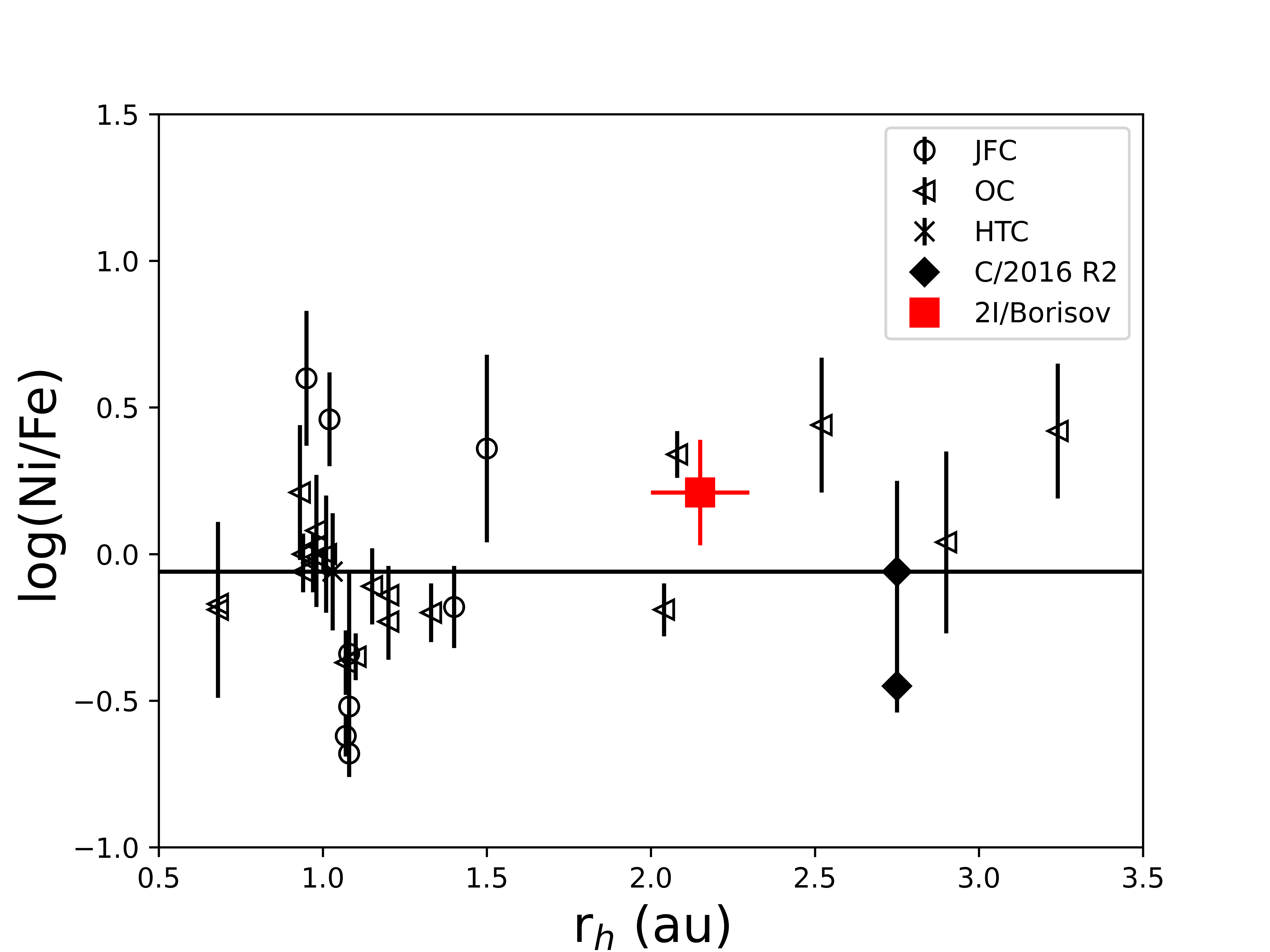}
  \caption{Comparison between the log(Ni/Fe) ratio of 2I and the values measured by \cite{Manfroid2021} for a sample of solar system comets (Jupiter Family Comets (JFC), Oort Cloud Comets (OC), and Halley Type Comets (HTC)). The horizontal line represents the average value for solar system comets from \cite{Manfroid2021}.}
   \label{Plot_NiFe}
\end{figure}

Within the uncertainties, the log(NiI/FeI)=~$0.21\pm0.18$ derived for 2I is in agreement with the average value of the ratio log(NiI/FeI)= $-0.06\pm0.31$ reported by \cite{Manfroid2021} for a sample of 17 solar system comets (see Fig. \ref{Plot_NiFe}). We also note that \cite{Manfroid2021} found a correlation between the total Fe and Ni production rates and water and CO production rates for most solar system comets. We use our Q(H$_2$O) value of $2.2\times10^{26}$ molecules/s and a Q(CO) of $8\times10^{26}$ molecules/s (interpolating between values for dates before and after our measurement from \cite{Bodewits2020}) to estimate how 2I fits into these trends. While it fits within the trend for the Q(CO) vs Q(Fe+Ni), 2I is an outlier for the Q(H$_2$O) vs Q(Fe+Ni), as illustrated in Fig. \ref{Plot_NiFe_H2O} and \ref{Plot_NiFe_CO}. Comet C/2016 R2 (PANSTARRS) is a solar system comet extremely rich in CO \citep{Biver2018}, like 2I. Similar to C/2016 R2, 2I fits within the Q(CO) vs Q(Fe+Ni) trend for solar system comets but not in the Q(H$_2$O) vs Q(Fe+Ni) trend, strengthening the similarity between 2I and C/2016 R2.


\subsection{NH$_2$ and NH$_3$ ortho-to-para ratios}

NH$_2$ lines were detected in most of our spectra and used to measure the ortho-to-para abundance ratio (OPR) of NH$_2$. The measurement of the OPR is of great interest, and even more in the case of an interstellar comet, as the relative abundance of different nuclear-spin isomers and the spin temperature that can be derived from it has been considered as an indicator of the temperature conditions prevailing when the molecules were formed, although this interpretation has been challenged recently \citep{Hama2011}. 

\begin{figure*}
\centering
\includegraphics[width=18cm]{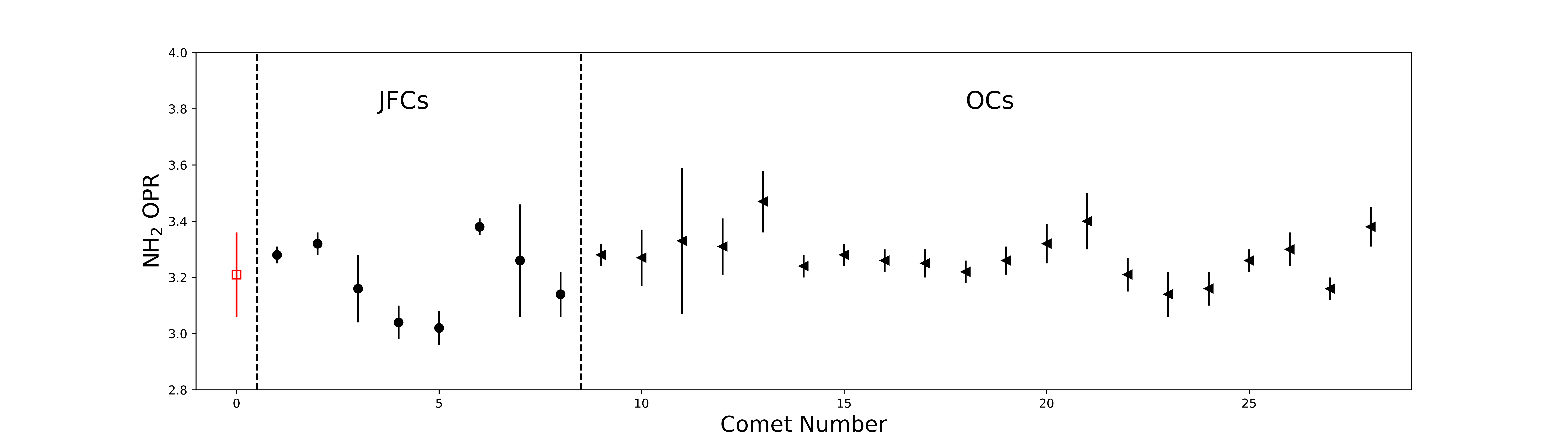}
  \caption{NH$_2$ OPR of 28 Jupiter Family (JFCs) and Oort Cloud (OCs) solar system comets (black symbols), compared to the value measured for the interstellar comet 2I (red square). Adapted from \cite{Shinnaka2016b}, with additional data from \cite{Yang2019,Shinnaka2020}.}
   \label{OPR_Graph}
\end{figure*}

\begin{table}[ht!]
\caption{\label{TableOPR} NH$_2$ and NH$_3$ ortho-to-para ratios and spin temperatures measured in the coma of 2I}
\centering
\begin{tabular}{llll}
\hline\hline

Date & r$_h$ & NH$_2$ OPR & NH$_3$ \\
\hline
NH$_2$ (0,8,0) band\\
\hline
2019-11-15 & 2.1 & $3.07\pm0.23$ & $1.04\pm0.12$ \\
2019-12-16 & 2.0 & $3.09\pm1.07$ & $1.05\pm0.54$ \\
2019-12-24 & 2.0 & $3.29\pm0.36$ & $1.15\pm0.18$ \\
2019-12-26 & 2.0 & $3.89\pm0.88$ & $1.45\pm0.44$ \\
\hline
Average &  & $3.17\pm0.19$ & \\
$T_\mathit{Spin}$ &  & $>27 (1\sigma)$ K & \\
$T_\mathit{Spin}$ &  & $>22 (3\sigma)$ K & \\
\hline
NH$_2$ (0,9,0) band\\
\hline
2019-11-15 & 2.1 & $3.63\pm0.38$ & $1.32\pm0.19$ \\
2019-12-16,24,26 & 2.0 & $2.99\pm0.39$ & $1.00\pm0.20$ \\
\hline
Average &  & $3.32\pm0.27$ & \\
$T_\mathit{Spin}$ &  & $28^{+16}_{-4}$ K  & \\
\hline
\end{tabular}
\end{table}

We measured the NH$_2$ OPR on 4 epochs in 2019, but the SNR was unfortunately not sufficient for the 2020 observations. We used the two ro-vibronic emissions bands (0,8,0) and (0,9,0), following the method described in \cite{Shinnaka2011}. For the (0,8,0) band we could determine the OPR separately for 2019 Nov 15, Dec 16, 24, and 26. For the (0,9,0) band, the Dec 16, 24, and 26 observations had to be averaged to measure the NH$_2$ OPR. The OPR of ammonia was computed considering NH$_3$ as the sole parent of NH$_2$, and keeping the total nuclear spin for the photodissociation reaction. Our measurements are listed for each band in Table \ref{TableOPR}. The weighted average NH$_2$ OPR value for both bands is 3.21 $\pm$ 0.15, corresponding to a NH$_3$ OPR of $1.11\pm0.08$. A nuclear spin temperature ($T_\mathit{Spin}$) for ammonia of $31^{+10}_{-5}$ K was derived. Within the uncertainties, we do not see an evolution of the OPR with time or heliocentric distance, consistent with what is usually observed for solar system comets.

2I's NH$_2$ OPR (and corresponding NH$_3$ OPR and spin temperature) are remarkably similar to the typical values measured in solar system comets (see Fig. \ref{OPR_Graph}). The ammonia OPR and spin temperature have been thought to be linked to the formation temperature of the molecule. If this is the case, the similar OPR and spin temperature values would indicated that 2I had a similar formation environment to our own solar system. However, recent laboratory experiments have shown that the water OPR might not retain the memory of the molecule formation temperature \citep{Hama2011,Hama2013,Hama2016} but rather be diagnostic of the physico-chemical conditions in the innermost coma or in the subsurface layers. If this is verified, a similar argument could be made for the ammonia OPR and spin temperature. It is thus likely that the similar NH$_2$ OPR for 2I and solar system comets simply reflect similar physico-chemical conditions in the inner coma.


\subsection{Water production rate}
\label{waterprod}

\begin{figure}
\centering
\includegraphics[width=9cm]{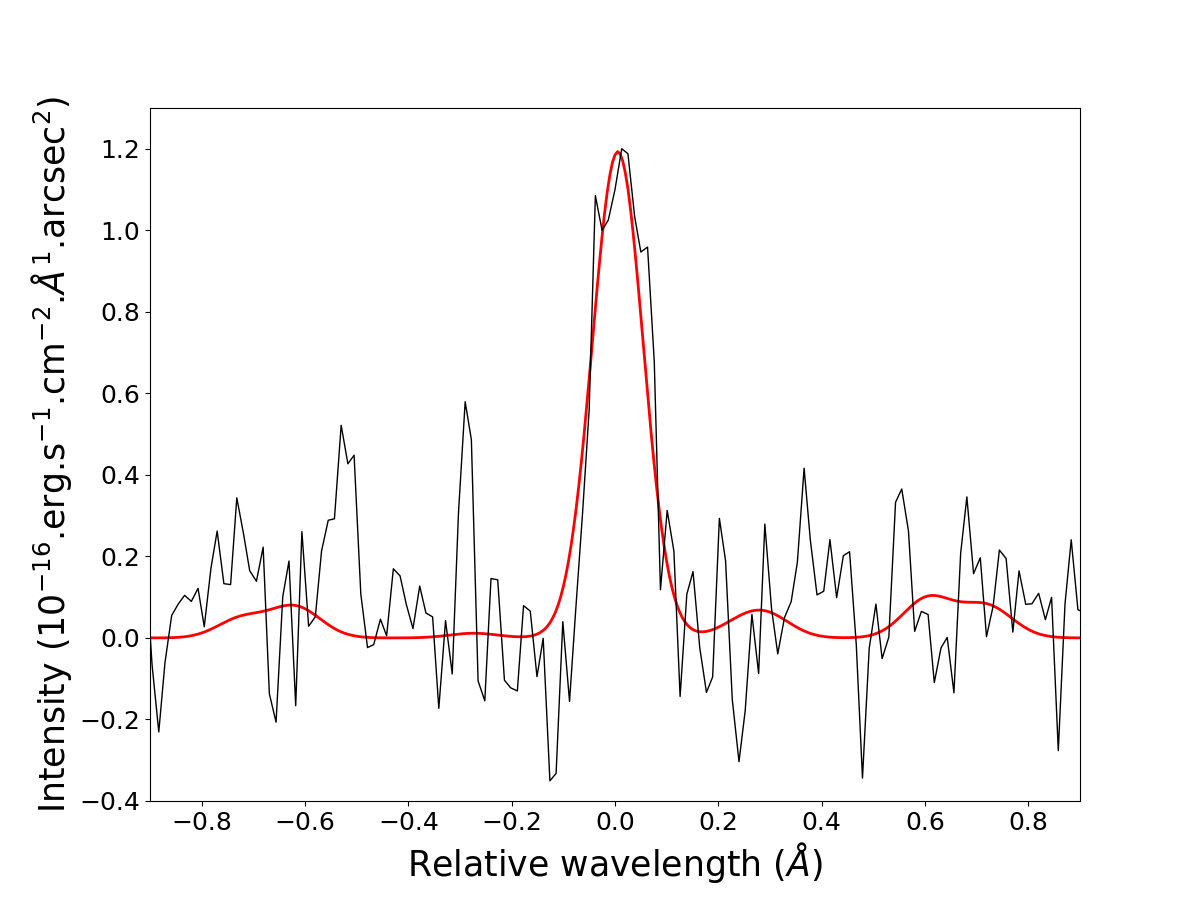}
  \caption{Average of the 17 brightest OH lines in the average spectrum of comet 2I obtained on Dec 24 and 26 with UVES at the VLT. The red line represents the fitted fluorescence model.}
   \label{Plot_OH}
\end{figure}

As mentioned before, we detected the OH (0-0) band around 309 nm in our spectra, from which we compute the production rate of water. We used spectra obtained on Dec 24 and 26, about two weeks after the comet's perihelion passage, as these were the ones with the best SNR. To further increase the SNR, we averaged the two spectra. We used a Haser model to compute the production rate, with scalelengths from \cite{Cochran1993} for OH, and our own model of OH fluorescence spectrum \citep{Rousselot2019} for computing the fluorescence efficiency. This model provided a fluorescence efficiency of $6.0\times 10^{-16}$~erg.s$^{-1}$.molecule$^{-1}$ with the average heliocentric distance and velocity of 2I corresponding to the two spectra (2.04~au and +7.02~km.s$^{-1}$, respectively). Such a value can be compared, e.g. to the fluorescence model published by \cite{Schleicher1988} that would provide about $6.5\times 10^{-16}$~erg.s$^{-1}$.molecule$^{-1}$ for similar parameters. This small difference could be due to different parameters used for both model (e.g., transition probabilities, solar atlas). We assumed an equal velocity for the daughter and parent of 595~m/s, following the formula v(H$_2$O) = $850/\sqrt(r_h)$~m/s given in \cite{Cochran1993}. Given the low SNR around 309~nm, we summed the seventeen brightest apparent lines (corresponding to 27 real lines, some of them being blended) and measured the corresponding total flux equal to $4.5\pm0.5 \times10^{-15}$~erg.s$^{-1}$.cm$^{-2}$ for the 1.8\arcsec-wide slit (see Fig. \ref{Plot_OH}). The final total intensity is sensitive to different parameters, such as errors in offset and baseline corrections. The final spectrum allows us to estimate these errors to about $0.5 \times10^{-15}$~erg.s$^{-1}$.cm$^{-2}$, with the average baseline being very close to zero in intensity. With the parameters given above, we computed a production rate Q(OH) = $1.9\pm0.2 \times 10^{26}$~molecules/s. Assuming a ratio of the production rate of OH radicals to the total photodestruction rate of water (photodissociation plus ionization) for  a quiet Sun (corresponding to the time of observations) of
 87\% \citep{Crovisier1989} this corresponds to a water production rate of  Q(H$_2$O) = $2.2\pm0.2 \times 10^{26}$~molecules/s. 
 
 A water production rate from UVES spectra was already reported by \cite{Jehin2020b}, and is consistent with what we find in this work after improving the flux calibration. Water production rates were also measured by \cite{Xing2020} between Nov 1 and Dec 21, 2019.  On Dec 21, they measure Q(H$_2$O) = $4.9\pm0.9\times 10^{26}$~molecules/s. This is more than two times higher than the value we measured only a few days later. However, we note that \cite{Xing2020} use a vectorial model to derive the water production rate, while we use a Haser model. Re-computing the water production rate with the same model and model parameters, we find Q(H$_2$O)=$3.4 \times 10^{26}$~molecules/s. Considering the steep post-perihelion decrease in the water production rate reported by \cite{Xing2020}, this value is consistent with their measurements and confirms the decreasing trend.

\subsection{Forbidden oxygen emission lines}
As shown in Table \ref{TableSpec}, we detected forbidden oxygen lines at 557.73, 630.03, and 636.38~nm in the coma of 2I. Atomic oxygen atoms in the coma of comets are produced by the photo-dissociation of oxygen-bearing species, the most abundant contributors being H$_2$O, CO, CO$_2$, and O$_2$. It has been shown that the ratio between the intensity of the green oxygen line (at 557.73~nm) and the red-doublet (at 630.03 and 636.38~nm) depends on the relative abundance of the different parents in the coma \citep{Festou1981,McKay2013,Decock2013,Decock2015}. This ratio is usually referred to as the G/R ratio. The cometary forbidden oxygen lines could only be resolved from their telluric counterparts and the G/R ratio measured from some spectra.
On Nov 15, 2019 we measured a G/R= $0.31\pm0.05$, while the comet was at 2.1 au from the Sun pre-perihelion. We combined the spectra from Jan 29, Feb 1, and Feb 2, 2020 to increase the signal-to-noise ratio and found a G/R= $0.3\pm0.1$ (at r$_h$=2.3 au). We also combined the data from Feb 22 and Mar 16 to better constrain its evolution with time and heliocentric distance and measured G/R= $0.6\pm0.3$. More details about the measurement of the G/R ratio can be found in appendix \ref{App_GR}.

\begin{figure}
\centering
\includegraphics[width=9cm]{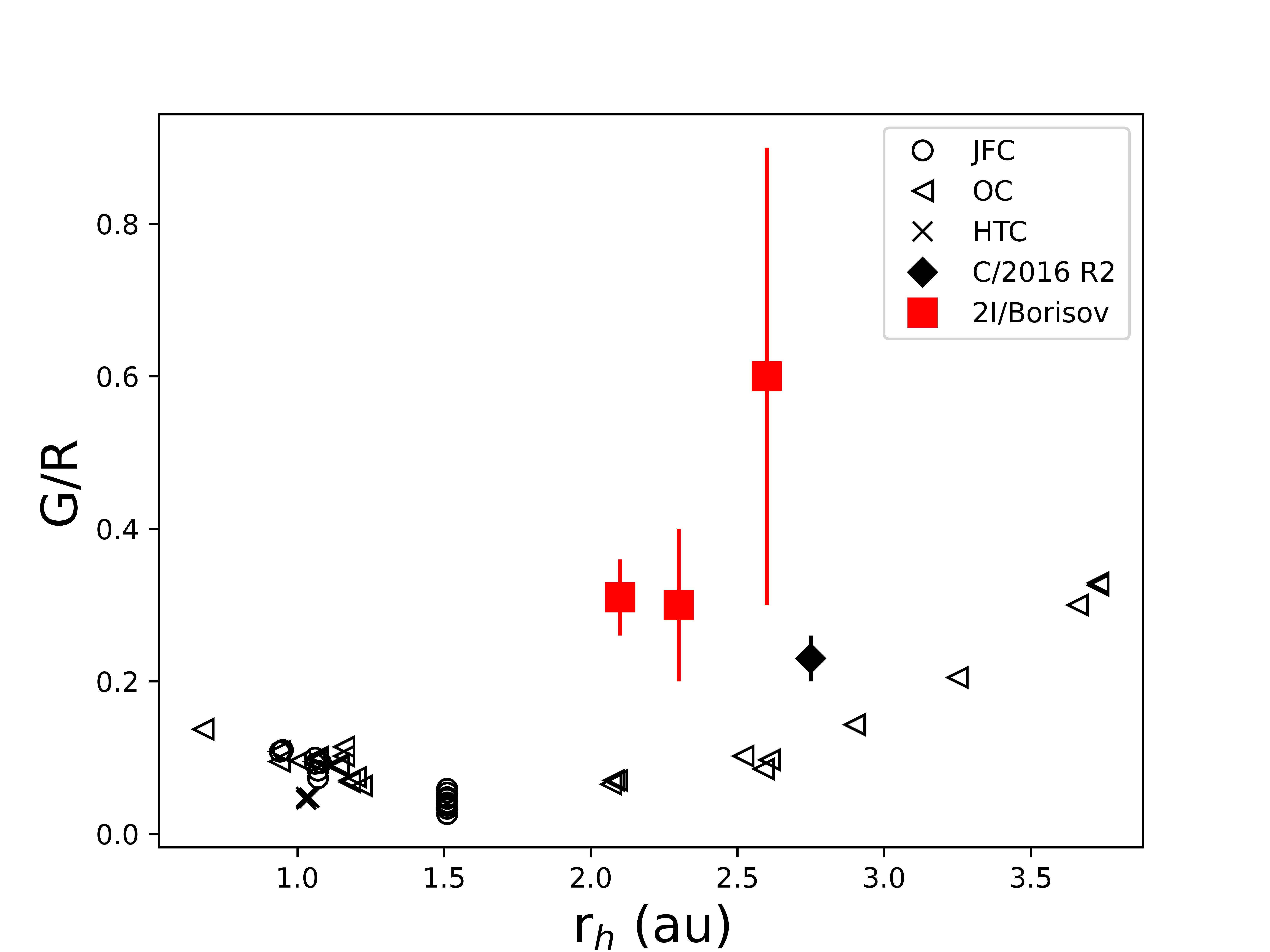}
  \caption{Ratio between the green and the sum of the two red forbidden oxygen lines (G/R) in 2I compared to that measured for a sample of solar system comets by \cite{Decock2013} (and \cite{Opitom2019b} for C/2016 R2 - Jupiter Family Comets (JFC), Oort Cloud Comets (OC), and Halley Types Comets (HTC)).}
   \label{Plot_GR}
\end{figure}

These values are high compared to what is usually measured in comets at the same distance from the Sun, as illustrated in Fig. \ref{Plot_GR}. For a sample of 11 comets, \cite{Decock2013} report a mean value of $0.11\pm0.07$. It has been observed by both \cite{Decock2013} and \cite{McKay2013} that the values of the G/R ratio increases for comets observed at large heliocentric distances, but this increase of the G/R ratio is usually noticed for comets beyond 2.5 au \citep{Decock2013}. The only other outlier in Fig. \ref{Plot_GR} is C/2016 R2, which like 2I was very rich in CO. The G/R ratio expected if the oxygen atoms are produced from the photo-dissociation of water is around 0.08  but is much higher (around 0.78) if they are produced by the photo-dissociation of CO \citep{Bhardwaj2012}. The higher than usual G/R ratio we measure for 2I might thus be the consequence of its high CO/H$_2$O ratio. As the heliocentric distance increases, we would expect the decrease in the water production rate to cause a decrease of the intensity of the red-doublet. At the same time, the intensity of the green line, to which the CO contributes significantly, would not decrease as much. As a consequence, the G/R ratio would increase, which could explain the higher G/R ratio measured in March. Dedicated chemistry-emission modelling is out of the scope of this work but will be explored in the future to further investigate the cause of the high G/R ratio of 2I.

\section{Conclusions and perspectives}
We observed the first active interstellar comet 2I/Borisov with the UVES high resolution spectrograph on 12 dates between Nov 15, 2019 and Mar 16, 2020. These observations constitute a unique data set to constrain the composition of the interstellar visitor. Using these spectra, we performed various measurements, some being made for the first time in an interstellar comet.  

In particular, we measured a NiI to FeI abundance ratio consistent with what has been found for a sample of 17 solar system comets by \cite{Manfroid2021}. 
In the future, the Ni/Fe ratio measured in the coma of interstellar comets could complement molecular abundances already used and provide an additional tool to probe the composition of interstellar comets' native systems. Indeed, the relative abundances of metals produced by type 1a supernovae nucleosynthesis varies depending on the explosion mechanism or the initial conditions of the white dwarf \citep{Leung2018,Palla2021}. The material produced by those supernovae then feeds the interstellar medium and can be integrated in forming planetary systems, leaving an imprint on the composition of comets that could be measurable given the large variations of the Ni/Fe ratios observed in supernovae \citep{Mori2018}.

We also derived the NH$_2$ OPR (and corresponding NH$_3$ OPR and spin temperature) of 2I and find them similar to solar system comets. This likely reflects similar physico-chemical conditions in the inner coma. We detected forbidden oxygen lines and computed G/R ratios higher than what is usually found for solar system comets at that distance from the Sun. However, this is consistent with the high CO abundance in the coma of 2I and the increase of the CO/H$_2$O ratio with the heliocentric distance. In conclusion, our high spectral resolution observations of 2I/Borisov reveal a remarkable similarity to solar system comets in terms of OPR and Ni/Fe abundances. The G/R ratio is high, similar to C/2016 R2, confirming a high CO/H$_2$O abundance ratio, and suggesting that both comets might have formed in colder environments.

\begin{acknowledgements}
Based on observations collected at the European Southern Observatory under ESO programmes 2103.C-5068(F) and 105.205Q.006. We are grateful to the ESO Paranal staff and user support department for their efforts and exceptional support in obtaining the observations. CO is a Royal Astronomical Society Norman Lockyer fellow and University of Edinburgh Chancellor's fellow. JM, DH, and EJ are honorary Research Director, Research Director and Senior Research Associate at the F.R.S-FNRS, respectively.
\end{acknowledgements}

%

  \bibliographystyle{aa} 
   \bibliography{Biblio} 

%
\begin{appendix} 

\section{Detected species}

\begin{figure}[h!]
\centering
\includegraphics[width=9cm]{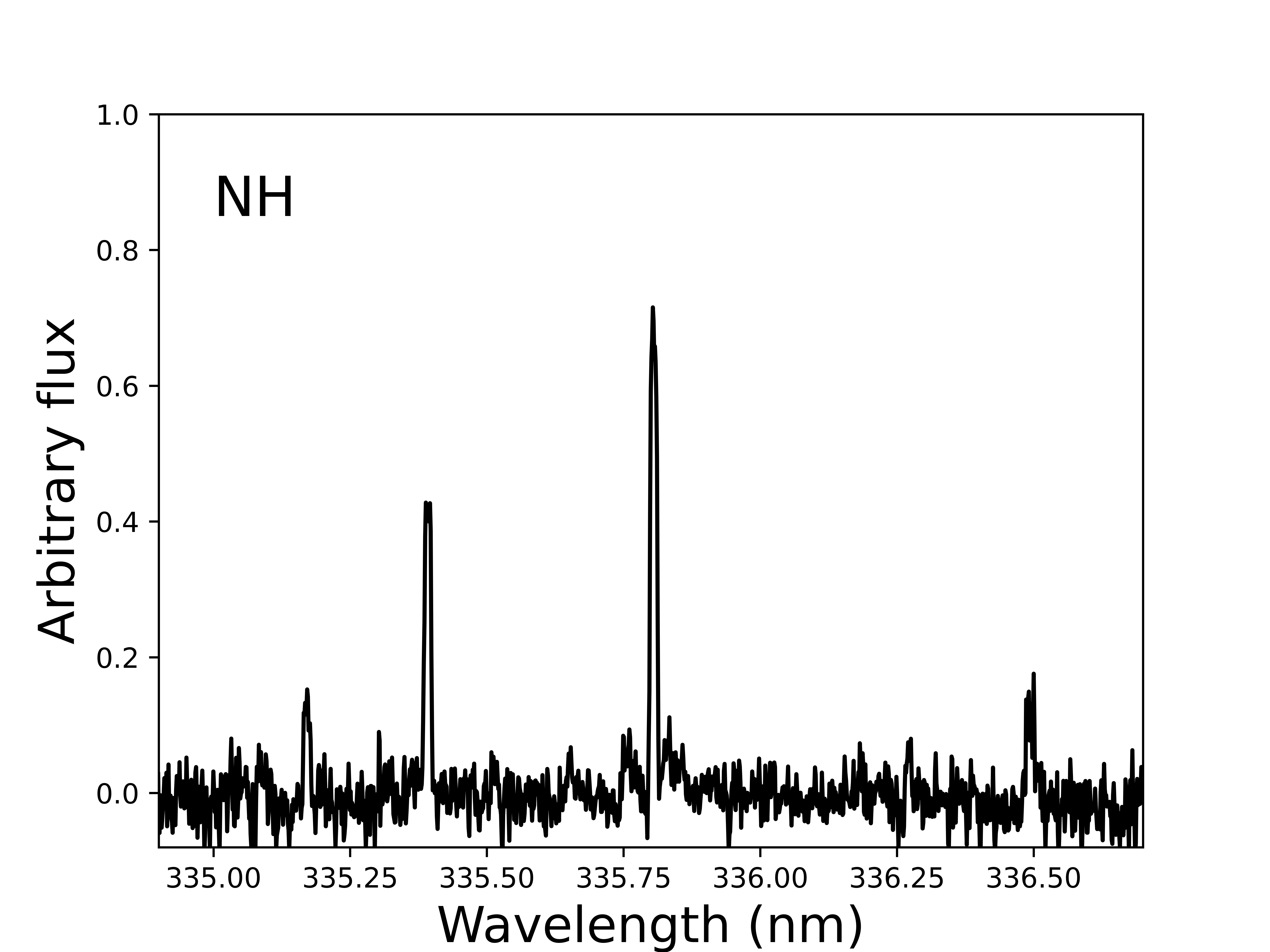}
  \caption{NH $\mathrm{A^3\Pi_i - X^3\Sigma^-}$ band in the coma of 2I/Borisov around 335 nm.}
   \label{Plot_NHLines}
\end{figure}

\begin{figure}[h!]
\centering
\includegraphics[width=9cm]{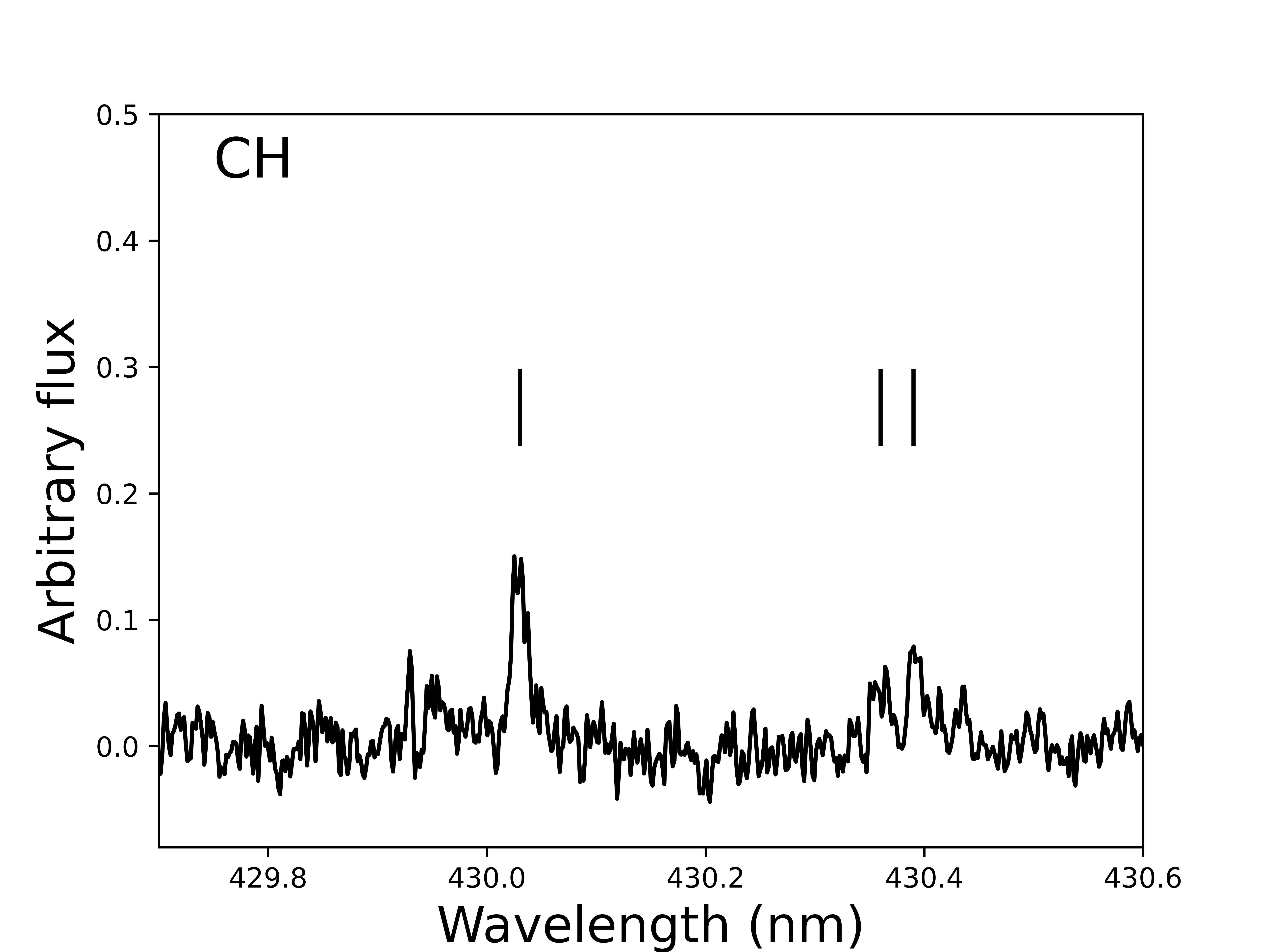}
  \caption{Spectrum of 2I over the 429.7-430.6 nm range covering 3 prominent CH lines from the (0,0) $\mathrm{A^2\Delta-X^2\Pi}$ band (indicated by vertical black lines).}
   \label{Plot_CHLines}
\end{figure}

\begin{figure}[h!]
\centering
\includegraphics[width=9cm]{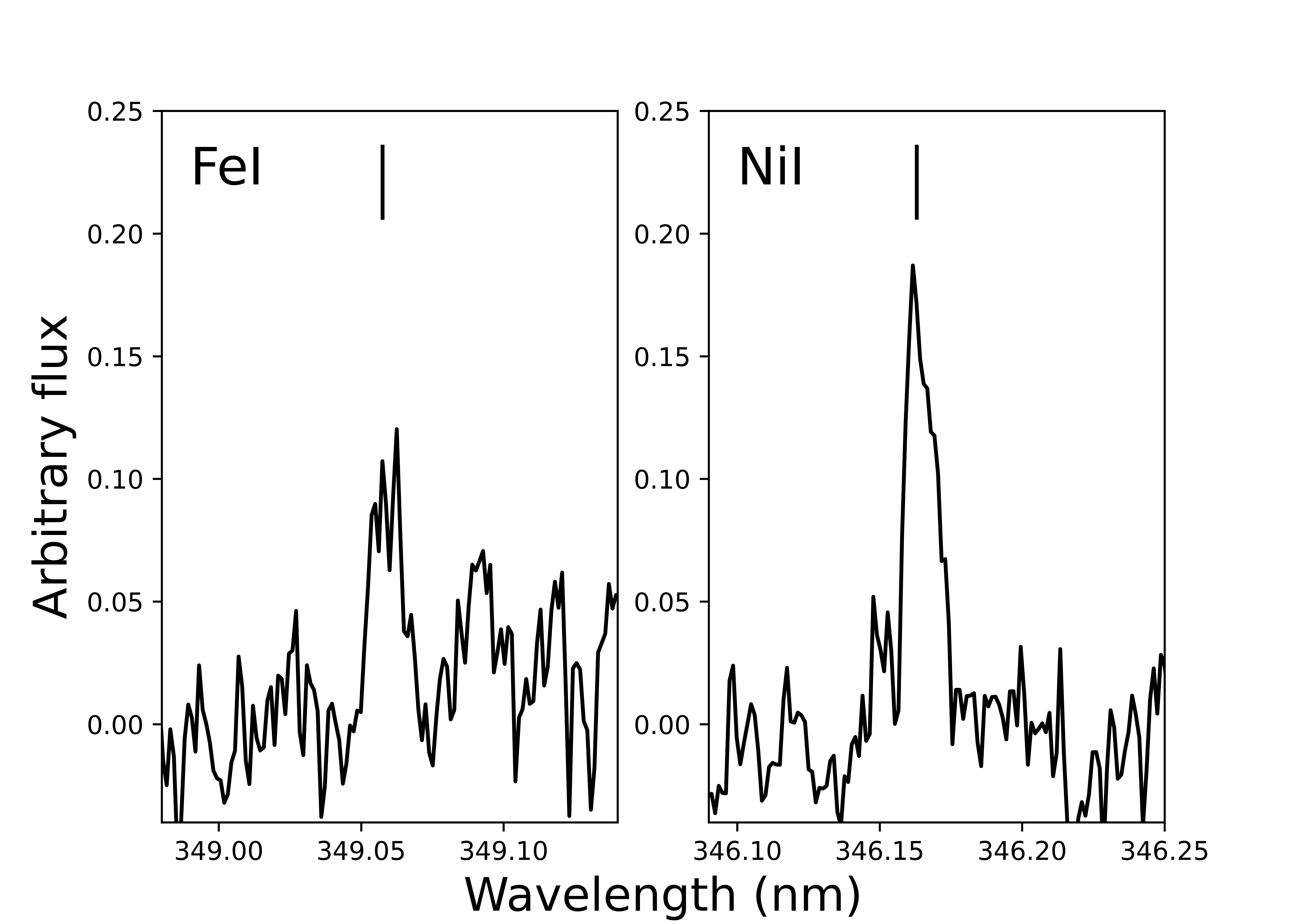}
  \caption{Example of FeI and NiI lines detected in the coma of 2I at 349.06 and 346.17 nm, respectively.}
   \label{Plot_NiFeLines}
\end{figure}

In this appendix, we give the details of the species and specific emission bands detected in our high resolution spectra of 2I. Table \ref{TableSpec} summarizes our detections and non-detections. In addition to the NH, CN, C$_2$, and NH$_2$ emission bands, the 630~nm [OI] line, and neutral nickel emission lines that were detected in previous observations \citep{Fitzsimmons2019,Bannister2020,Aravind2021,Kareta2020,McKay2020,Guzik2021}, we report the detection of a CH emission band, 557.7 and 636.4~nm forbidden oxygen lines, as well as emissions lines from FeI. Our detection of NH, CH, FeI, and NiI emissions are illustrated in Figures \ref{Plot_NHLines}, \ref{Plot_CHLines}, and \ref{Plot_NiFeLines}, respectively.

\begin{table*}[hb!]
\caption{\label{TableSpec} Species detected in the coma of 2I from UVES observations}
\begin{tabular}{ll}
\hline\hline
Species     & Detected band/lines \\
\hline
OH & (0,0) [310 nm] $\mathrm{A^2\Sigma^+-X^2\Pi}$ \\
NH &  $\mathrm{A^3\Pi_i - X^3\Sigma^-}$ [336 nm]\\
CN & (0,0) [388 nm], (1,0) [422 nm] $\mathrm{B^2\Sigma^+-X^2\Sigma^+}$\\
CH & (0,0) [431 nm] $\mathrm{A^2\Delta-X^2\Pi}$ \\
C$_2$ & (0,0) [517 nm] $\mathrm{d^3\Pi_g-A^3\Pi_u}$ Swan System  \\
NH$_2$ & (0,12,0) [520 nm], (1,7,0) [540 nm], (0,11,0) [543 nm], (0,10,0 )[570 nm], (0,9,0) [600 nm], (0,8,0) [630 nm], \\
 & (0,7,0) [665 nm], (0,6,0) [695 nm] A–X \\
$[$OI$]$ &  557.73, 630.03, and 636.37 nm \\
NiI & 336.96, 339.30, 341.48, 343.36, 344.63$^*$, 345.85, 346.17, 349.30$^*$, 351.03, 351.51, 352.45, 356.64$^*$, 361.94 nm \\
FeI & 344.06,349.06, 358.12, 371.99, 373.49$^*$, 374.56, 374.59, 374.83, 374.95, 375.82 nm\\
\hline
\end{tabular}
\tablefoottext{1}{The blended FeI and NiI lines are marked by *}
\tablefoottext{2}{Non detected species: N$_2^+$, CO$^+$, CO$_2^+$, H$_2$O$^+$, C$_3$.}
\tablefoottext{3}{The number between brackets represent the approximate wavelength of the emission bands}
\end{table*}

\section{Correlation between FeI, NiI, H$_2$O, and CO}
\label{Correl}

\begin{figure}[h!]
\centering
\includegraphics[width=8.0cm]{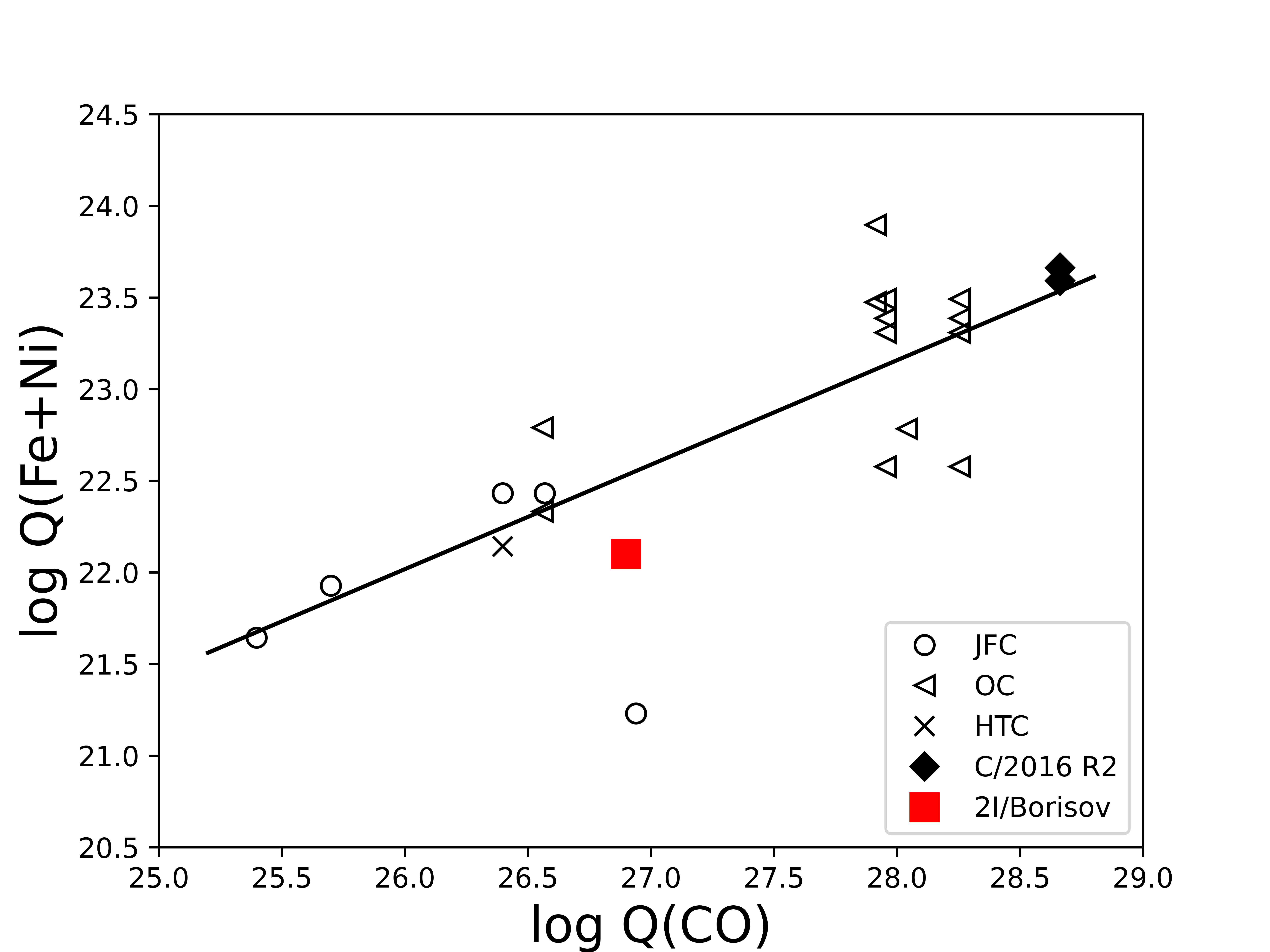}
  \caption{Sum of the production rates of FeI and NiI as a function of the CO production rate for 2I/Borisov, compared to the values measured by \cite{Manfroid2021} for a sample of solar system comets (Jupiter Family Comets (JFC), Oort Cloud Comets (OC), and Halley Type Comets (HTC)). The black line represents the best correlation for solar system comets from \cite{Manfroid2021} (excluding C/2016 R2). The JFC located below the trend is 9P/Tempel 1.}
   \label{Plot_NiFe_CO}
\end{figure}

\begin{figure}[b]
\centering
\includegraphics[width=8.0cm]{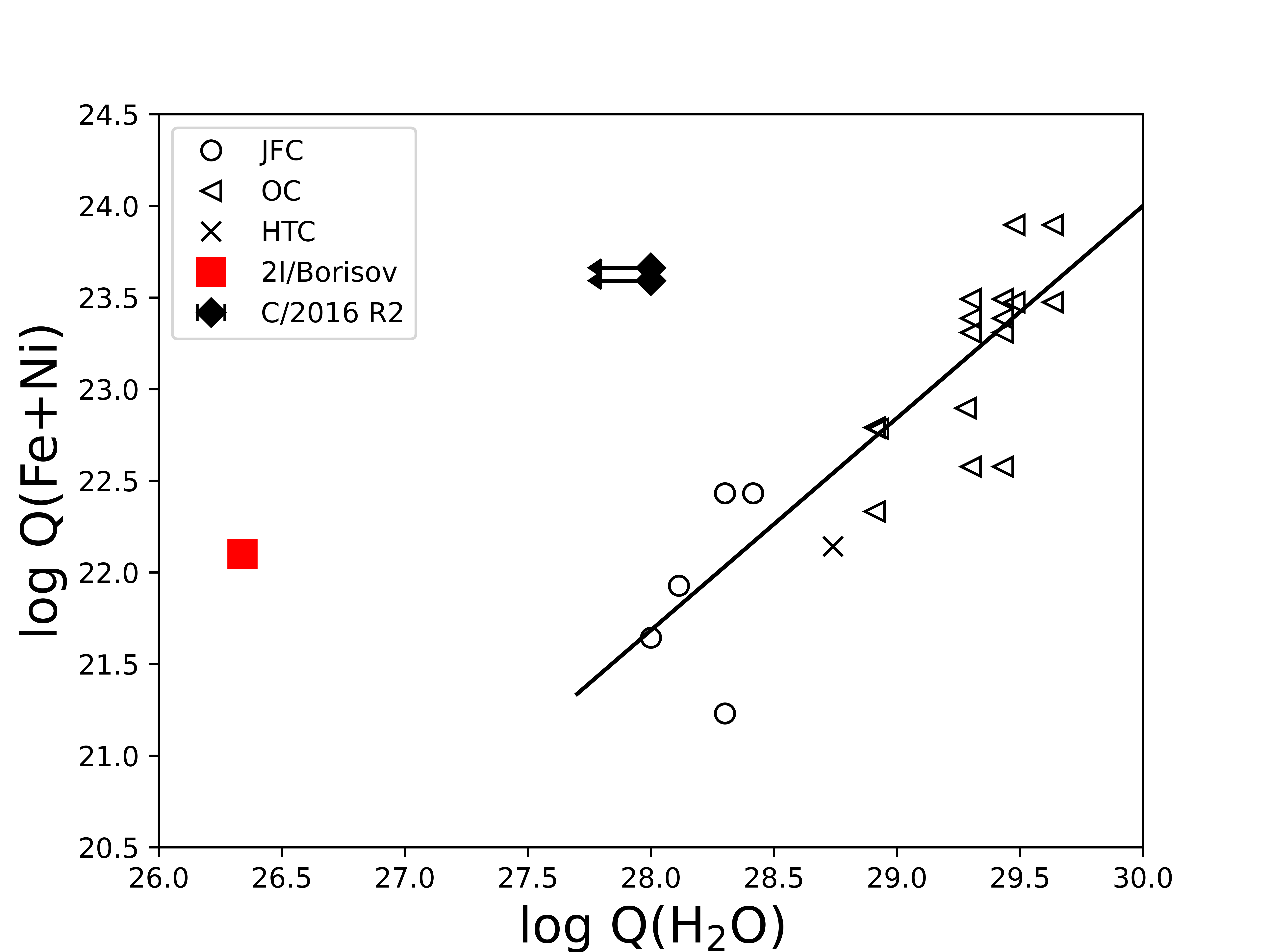}
  \caption{Sum of the production rates of FeI and NiI as a function of the water production rate for 2I/Borisov, compared to the values measured by \cite{Manfroid2021} for a sample of solar system comets (Jupiter Family Comets (JFC), Oort Cloud Comets (OC), and Halley Type Comets (HTC)).  The black line represents the best correlation for solar system comets from \cite{Manfroid2021} (excluding C/2016 R2).}
   \label{Plot_NiFe_H2O}
\end{figure}

In section \ref{FeNi}, we discuss correlations between the total Fe and Ni production rate with the CO and the water production rates in 2I compared to solar system comets measured by \cite{Manfroid2021}. In Fig. \ref{Plot_NiFe_CO}, which represents the total FeI+NiI production rate as a function of the CO production rate, 2I fits well within the trend defined by solar system comets. In Fig. \ref{Plot_NiFe_H2O}, however, 2I is outside the trend, indicating a lower water production rate compared to the total Fe+Ni production. The only other outlier is C/2016 R2 (PANSTARRS), which also does not fit with the trend defined by other solar system comets, reinforcing the similarities between 2I and C/2016 R2.

\label{Correl}

\section{Determination of the G/R ratio}
\label{App_GR}
\begin{figure}
\centering
     \begin{subfigure}{.5\textwidth}
     \centering
         \includegraphics[width=.99\linewidth]{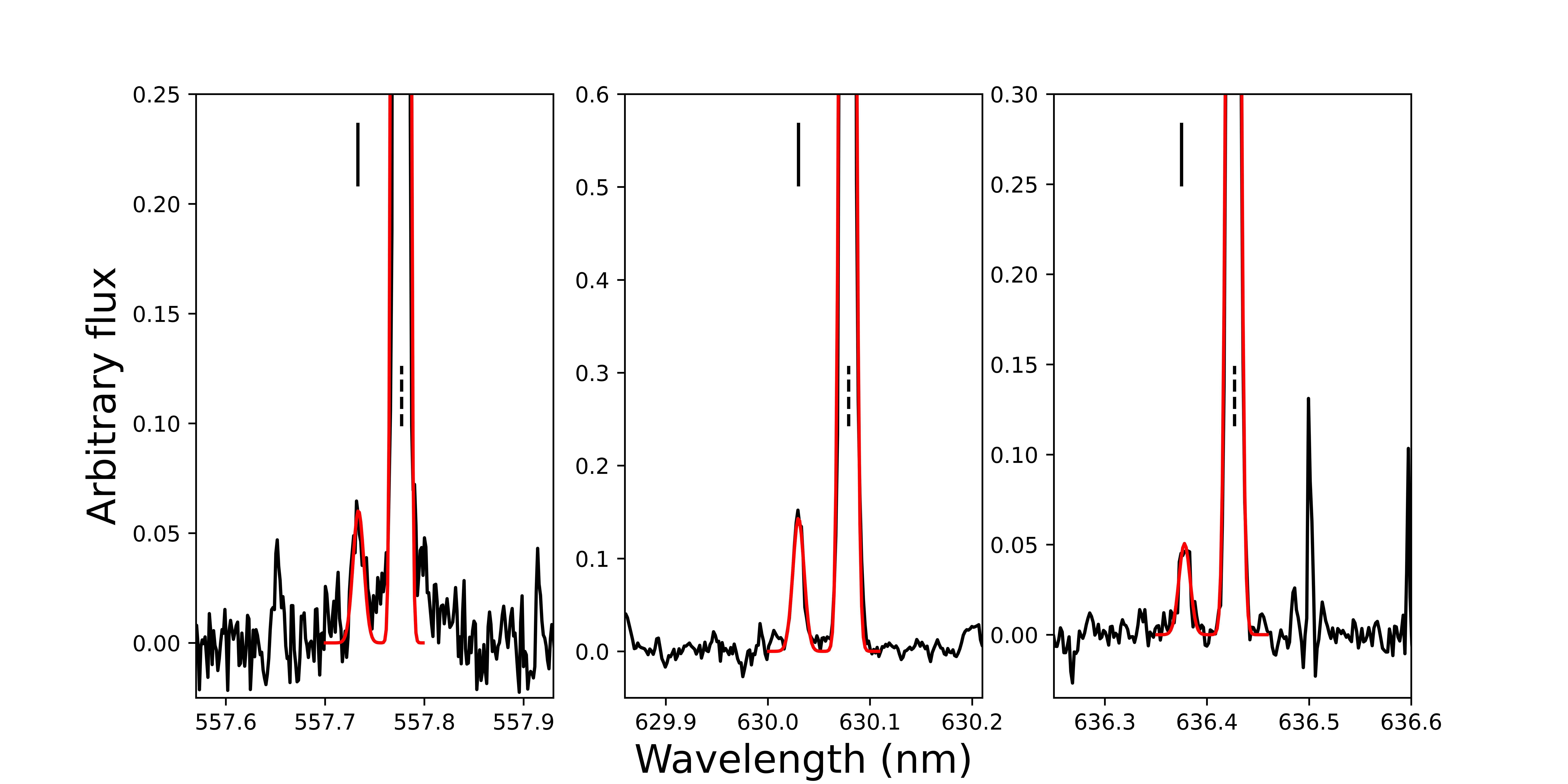}
         \caption{November 15, 2019}
         \label{GR_Nov}
     \end{subfigure}
     \newline
     \begin{subfigure}{.5\textwidth}
          \centering
         \includegraphics[width=.99\linewidth]{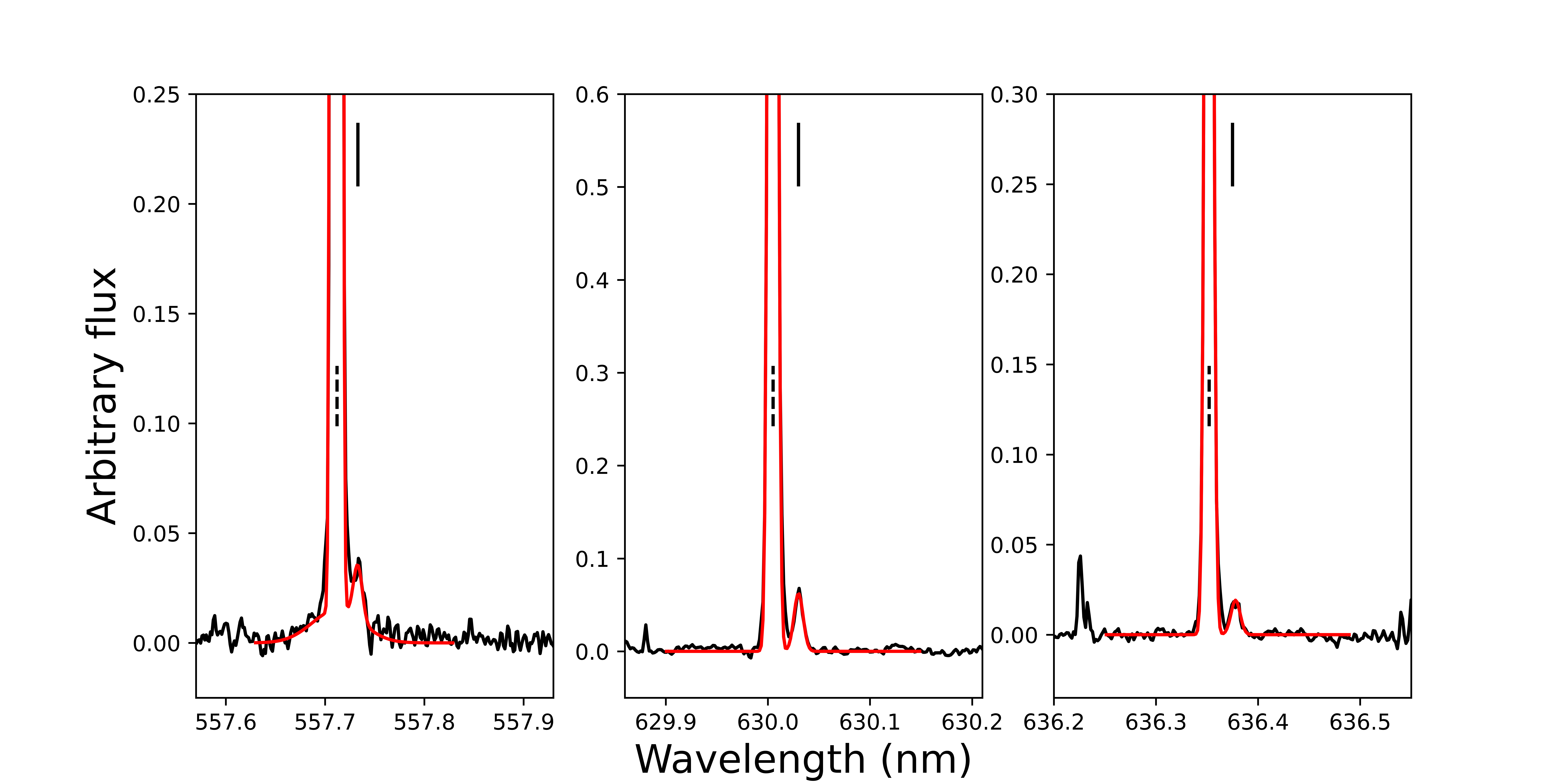}
         \caption{January 29, February 1, and February 2, 2020}
         \label{GR_Feb}
     \end{subfigure}
     \newline
     \begin{subfigure}{.5\textwidth}
          \centering
         \includegraphics[width=.99\linewidth]{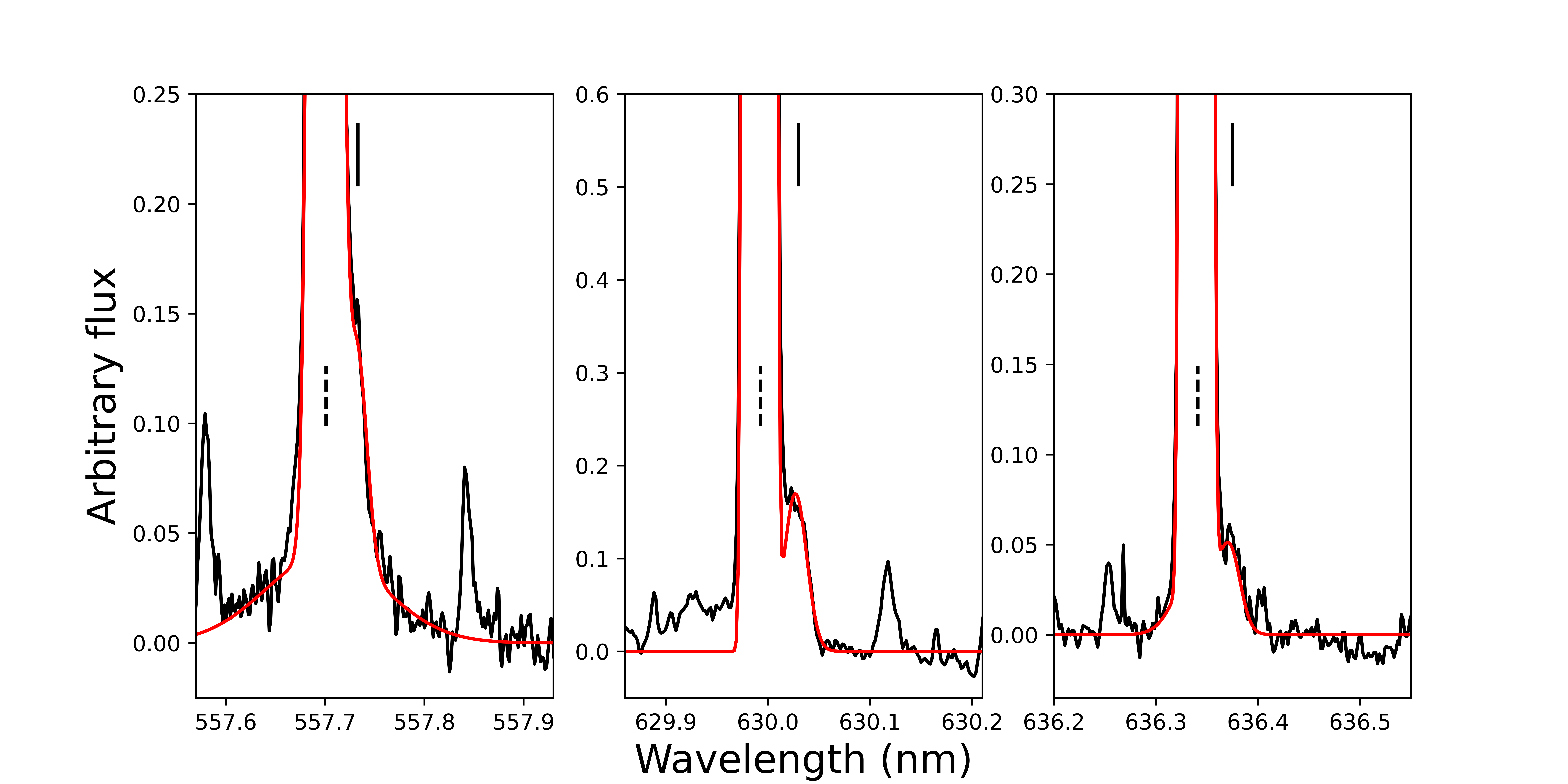}
         \caption{February 22 and March 16, 2020}
         \label{GR_Mar}
     \end{subfigure}
        \caption{Spectra of 2I around the 557.73 (left), 630.03 (centre), and 636.37 nm (right) forbidden oxygen lines at three different epochs (black line). Our best fit model, used to measure the G/R ratio is represented in red. The black vertical mark indicates the position of the cometary line and the dashed mark the position of the telluric line.}
        \label{GR_All}
\end{figure}

We were able to measure the G/R ratio at three different epochs. For each epoch, we modelled the cometary and telluric components of the forbidden oxygen lines simultaneously. For Nov 15, 2019, the 0.7\arcsec{} slit combined with the comet geocentric velocity of -22.8 km/s allowed us to resolve the cometary and telluric components. We used a single Gaussian to model each line and our best fit is shown in Fig. \ref{GR_Nov}. We then integrated the flux for the best-fit model of the cometary lines and computed the ratio between the intensity of the green line at 557.73 nm and the sum of the two red ones at 630.03 and 636.37 nm to obtain a G/R=$0.31\pm0.05$. For the first two epochs, we determined that the main source of uncertainty in the measurement of the G/R ratio was the subtraction of the dust continuum. We thus estimated the uncertainties by varying the level of the continuum at the wavelength of the oxygen lines prior to the subtraction by $\pm10\%$.

For the second epoch, we averaged 3 spectra from Jan 29, Feb 1, and Feb 2 2020, all obtained with a 0.4\arcsec{} slit. The geocentric velocity of the comet at that time was lower, which resulted in a minor blend of the telluric and cometary lines, especially for the green line. As can be seen in Fig. \ref{GR_Feb}, the 557 nm telluric line (left panel) has wide wings that are affecting the measurement of the cometary line. We thus used a sum of two Gaussians to represent the telluric [OI] line around 557.73 nm. All other lines could be satisfactorily reproduced by simple Gaussians. Due to the blend, we could not fit the width of the green cometary line and fixed it as the same as the red 630.03 nm line. Tests varying the width of the green line confirmed that the change in the  value of the G/R ratio is within the uncertainty of the measurement. The best fit is represented in Fig. \ref{GR_Feb} and resulted in G/R=$0.3\pm0.1$. 

For the last epoch, we used spectra from Feb 22 and Mar 16, obtained with a wide slit of 1.8\arcsec{}, which resulted in a blend of the telluric and cometary lines. Fortunately, even for the green line, we still see a knee in the telluric line corresponding to the comet emission. Because such a large slit was used, the telluric lines were not well reproduced by a Gaussian. To better match the shape of the telluric lines, we convolved the Gaussian with a box function. We added one Gaussian for the 636 nm line and two for the 557 nm line to satisfactorily reproduce the wings of the telluric lines. In this case, we had to manually vary the fit parameters to reach the best fit shown in Fig. \ref{GR_Mar}, corresponding to G/R=$0.6\pm0.3$. We estimated the uncertainty by varying the parameters of the model and computing the difference in the G/R ratio value between the most extreme fits that provided a satisfactory representation of the data. Because of the difficulty of de-blending the cometary and telluric line and the high uncertainty it causes on the measurement of the G/R ratio, we consider this measurement as tentative.

\end{appendix}

\end{document}